\documentclass[aps,draft,epsf,twocolumn]{revtex4}
\renewcommand{\v}[1]{{\bf #1}}

\newcommand{\gr}{{\nabla}}
\def\eqa{\begin{eqnarray}}
\def\eea{\end{eqnarray}}
\newcommand{\eq}{\begin{equation}}
\newcommand{\ee}{\end{equation}}
\newcommand{\nn}{\nonumber\\}
\newcommand{\Eq}[1]{Eq.~(\ref{#1})}

\newcommand{\ra}{\rightarrow}

\begin{document}
\title{The types of Mott insulator}
\author{Dung-Hai Lee$^{a,b}$ and Steven A. Kivelson$^{c}$}

\address{${(a)}$Department of Physics,University of California,
Berkeley,  CA 94720, USA\\ ${(b)}$Center for Advanced Study,
Tsinghua University, Beijing 100084, China\\ ${(c)}$Physics
Department,University of California, Los Angeles, CA 90095-9547
USA}
\begin{abstract}
There are two classes of Mott
insulators in nature, distinguished by their responses to weak
doping. With increasing chemical potential, Type I Mott insulators
undergo a first order phase transition from the undoped to the
doped phase. In the presence of long-range Coulomb interactions,
this leads to an inhomogeneous state exhibiting ``micro-phase
separation.''  In contrast, in Type II Mott insulators charges go
in continuously above a critical chemical potential. We show that
if the insulating state has a broken symmetry, this increases the
likelihood that it will be Type I. There exists a close analogy
between these two types of Mott insulators and the familiar Type I
and Type II superconductors.
\end{abstract}
\pacs{PACS numbers: 74.25.Jb,74.25.-q,74.20.-z} \maketitle
\narrowtext


The study of Mott insulators and doped Mott insulators
constitutes one of the central focuses of condensed matter physics
of the past two decades.  At a simple level, a Mott insulator is a
system in which the strong repulsion between particles impedes
their flow; the simplest cartoon of this is a system with a
classical ground-state in which there is one particle on each site
of a crystalline lattice and such a large repulsion between two
particles on the same site that
fluctuations
involving the motion of a particle from one site to the next are
suppressed.   The Mott insulator is thus essentially
classical in nature (and so accessible to particles of
any\cite{anyon} statistics.) In contrast a band
insulator\cite{band}, including all familiar semiconductors,  is a
state produced by subtle quantum interference effects which arise
from the fact that electrons are fermions. Nevertheless, one generally
considers band insulators to be ``simple" because the band theory
of solids successfully accounts for their properties;
although the physical origin of the insulating character of a Mott
insulator is understandable to any child, other properties,
especially the response to doping, are still only at best partially
understood.

There is even debate concerning the
definition of a Mott
insulator.  In many of the most familiar cases\cite{eg} involving
electrons, one can, in theory, continuously vary the strength of
the repulsion between electrons, such that in the limit of
vanishing strength interactions the system can be well understood
on the basis of band theory, while in the strong interaction limit
Mott physics better accounts for the insulating properties;  if
there is no phase transition between these two limits, it is not
clear at what interaction strength one should start thinking of
the system as a Mott insulator. However, recently an interesting
Bose Mott insulator has been produced by trapping bosonic neutral
atoms in an optical lattice\cite{bosemott};  there is no
non-interacting limit in which bosons are insulating, so no such
ambiguity arises here.

Mott states, in addition to being insulating, can be characterized
by the presence or absence of a spontaneously broken symmetry
(e.g. spin antiferromagnetism), by the nature of the low energy
excitation spectrum (e.g. gapped or gapless)\cite{note}, and by
the presence or absence of topological order and charge
fractionalization\cite{krs,rk,bfg,sondhi}.  To this list, we add a
``type'' index that classifies Mott insulators into two categories
depending on their response to doping.

Generally speaking, states with charge gaps (this includes both
band and Mott insulators) occur in crystalline systems at isolated
rational ``occupation numbers,'' $\nu=\nu^*$,  where $\nu$ is the
number of particles per chemical unit cell.  By ``doping'' we mean
a process which causes the occupation number to shift away from
$\nu^*$. When lattice translation symmetry is not spontaneously
broken, $\nu^*$ is typically an integer for bosons, consistent
with the simple classical cartoon described above, and an even
integer for spin 1/2 fermions (e.g. electrons).  The significance
of this latter statement is that a necessary condition for  an
electronic system to be adiabatically connected to a weakly
interacting band-insulator is that $\nu^*$ be an even integer.
Insulating states can also occur when $\nu^*$ is a
fraction (for fermions this includes odd-integer $\nu^*$). Usually
when that happens, translational symmetry is spontaneously broken
so that the new unit cell  has
integral occupation number (or even integral occupation number for
fermions)\cite{leesh,oshikawa}. For instance, electronic Mott
insulators with $\nu^*=1$ often exhibit antiferromagnetic N{\`e}el
long-range order, which doubles the unit cell leading to an
effective $\nu^*_{eff}=2$. Nevertheless, we will mention
\cite{rk,sondhi,bfg} model bosonic systems for which the Mott
state can be shown to have no broken symmetries for $\nu^*=1/2$,
and fermionic systems in which $\nu^*=1$. (Currently,  no
laboratory system has been found which unambiguously exhibits this
exotic behavior.)

The purpose of this paper is to address the response of Mott
insulators to light doping (i.e. $\nu\ra\nu^*-\delta$). Our
central observation is that there are two types of Mott insulators
(henceforth referred to as Type I and Type II). In a Type I Mott
insulator an increasing chemical potential induces a first order
phase transition from an undoped state to a charge-rich state so
that   changes occur discontinuously. In other words, there is a
range of ``forbidden charge density,'' and two-phase coexistence.
In a Type II Mott insulator, charges go in continuously above a
critical chemical potential. Since doping of a band insulator is
continuous, a doped Type II Mott insulator may be adiabatically
connected to a band insulator, but a doped Type I Mott insulator
is a thermodynamically distinct state of matter.

In the presence of long-range Coulomb forces macroscopic charge
inhomogeneity is impossible. However, so long as the long-range
forces are not too strong, there can exist a range of low doping
in which the doped charges tend to cluster\cite{low}. We still refer to this
kind of system as a Type I Mott insulator\cite{coulomb}. For
instance, if a low density of doped charges form puddles or
stripes\cite{andelman}, we consider this a form of charge clustering,
while a Wigner crystal of doped charges is
deemed homogeneous.

What is the situation in real materials?  That simple semiconductors can be continuously
doped, and hence are Type II, is well known, although it is not clear they are profitably
thought of as Mott insulators,  even when
correlation effects significantly renormalize the gap magnitude.  There
are many materials in which the insulating state is also
antiferromagnetically ordered.  It has been
argued\cite{nagaev,khomskii,ek} that doping of this class of Mott
insulators generically\cite{noteaf} leads to phase separation, {\it i.e.}
that they are Type I.  We will show below that the presence of a broken
symmetry always increases the tendency of an insulator to be Type I,
although the conjectured strict connection has neither been
proven\cite{pryadko} nor disproved.

For the cuprates and manganates, both of which have
antiferromagnetic order when undoped, there exists considerable
evidence that  doping induces spatial
inhomogeneity\cite{ref1,stm,stm1,ref2,ref3}. Conversely,
Sr$_{1-x}$La$_x$TiO$_3$, which is also an antiferromagnet when
undoped\cite{keimer}, is generally believed\cite{tokurarmp} to be
Type II on the basis of the observation of conventional Fermi
liquid behavior\cite{lsto} for doping as low as $x=5\%$, with an
effective mass that shows a tendency to diverge with decreasing
$x$;  taken at face value\cite{but}, this evidence of a quantum
critical precursor to a continuous metal-insulator transition
indeed suggests that this material is Type II.


Another purpose of this paper is to elucidate the
analogy between Type I and Type II Mott insulators
and the familiar types of superconductors. In particular, when the
constituent particles are bosons there exists a mathematically
precise mapping, the so-called ``duality
transformation''\cite{leesh,fl}, that relates the zero-temperature
properties of the doped Mott insulator in two spatial dimensions,
D=2,  to the finite-temperature response of a 3D superconductor to
a magnetic field. Table I summarizes the correspondence
between the two. This same mapping was exploited previously by Balents,
Fisher, and Nayak\cite{bfn} in the context of a theory of
a proposed  ``nodal liquid'' Mott insulating state;  in this regards, the present paper
simply underlines the generality of the analogy.
\begin{widetext}
\begin{tabular}{|l|l|} \hline
{\it $T=0$ properties of 2D Bose Mott insulators} & {\it $T>0$ properties of 3D Superconductors} \\
\hline\hline Doping & Applying magnetic field \\ \hline Chemical
potential $\mu$ & Applied magnetic field H\\ \hline Induced particle density $\rho$ & Magnetic induction $B$ \\
\hline World line of doped particles & Flux tubes\\ \hline Quantum
delocalization of doped particles & Thermal
meandering of flux tubes\\ \hline Type I Mott insulator & Type I superconductor \\
 Mott gap & $H_c$\\ Effective attraction between doped particles &
Positive N-S interface energy\\  \hline Type II Mott insulator &
Type II superconductor\\ Effective repulsion between doped
particles & Negative
N-S interface energy\\
 Mott gap & $H_{c1}$ \\ Wigner crystal of
doped particles &Abrikosov flux lattice \\
Superfluid state & Entangled vortex fluid\\
Critical $\mu$ at which Wigner crystal melts &$H_{c2}$\\
\hline
\end{tabular}
\vskip .1in
\end{widetext}

\section{Doping a translationally invariant Bose Mott insulator}

To begin with, we avoid the interesting complications which arise
from spontaneous symmetry breaking or from the fermionic character
of electrons by focusing on the simplest kind of Mott insulators
formed by spin zero point bosons on a lattice at  $\nu=1$. This is not a purely academic
exercise\cite{dope}, as it applies rather directly to the experiments in
Ref.\cite{bosemott}. Consider the
following Hamiltonian \eqa H&&=-{t\over 2}\sum_{<ij>} ( a_i^+a_j+
h.c.) + {U\over 2} \sum_i (a_i^+a_i)(a_i^+a_i-1)\nn&&+ {1\over
2}\sum_{i,j} V_{ij} a_i^+a_ia_j^+a_j\
-\mu\sum_ia^+_ia_i,\label{model} \eea where $i,j$ label the
lattice sites on a D-dimensional hypercubic lattice, and $a_i^+$
creates a boson at site $i$. The first term of \Eq{model}
describes the quantum mechanical ``hopping'' of bosons from a site
$i$ to its nearest
neighbors $j$, 
the second and  third terms  describe the pair-wise interactions
between bosons. The $U$ term is a contact interaction and the
$V_{ij}$ terms describe interaction between bosons separated by
$|\v r_i-\v r_j|$. This model is known as the extended
Bose-Hubbard model, and has been extensively studied from various
perspectives\cite{BH}.

Let us focus on the $U/t \to \infty$ limit.
In this ``hardcore'' limit,  doubly occupied sites cost  too much
energy, hence are excluded.
In this case for large positive $\mu$ there is an unique ground
state
\eq |{\rm Mott}>=\prod_j a_j^+|0>, \label{state} \ee in which each
site is occupied by one and only one boson and hence $\nu=1$.
Since this state is separated from all other states with the same
particle number by an energy gap of order $U$, clearly we have an
insulator.  (For $\mu$ large and negative, the groundstate is the
empty lattice, $|0>$.)

The behavior of the system upon varying $\mu$ ({\it i.e.} doping)
depends on the values of $V_{ij}$. For $V_{ij}=0$ the doped holes
(i.e the empty sites) interact only through the hardcore
exclusion. Since the kinetic energy favors a uniform
density state with delocalized holes, this limit corresponds to a
Type II Mott insulator. Indeed, a small concentration of doped
holes, $\delta\equiv 1-\nu<<1$, is equivalent to  a system of dilute
bosons (defined relative to the vacuum state $|Mott>$) with short-range interactions, a
problem whose solution has long been known\cite{Popov,fisherhohenberg}.  The energy is
clearly minimized by delocalizing the holes so the ground state in D = 2
is a uniform superfluid with superfluid stiffness proportional to
$t\delta$. Conversely, if $V_{ij}$  is attractive (negative),
it is clear that the holes will cluster when $|\sum_i V_{ij}|\gg
t$. Thus depending the nature of $V_{ij}$ \Eq{model} can describe
either a Type I or Type II Mott insulator. Note that in this example,
the Mott state does not break any symmetry, so any
inhomogeneity of the the doped state cannot be attributed to order
parameter competition.
The Type I behavior discussed here is more general\cite{beg} than that
which occurs in the bicritical point scenario in the Landau
theory of two competing order parameters\cite{zhang}.

To be still more explicit, consider the case in which $V_{ij} =-V$
for (ij) nearest-neighbor sites, and  $V_{ij} =0$ otherwise.
Technically in this limit \Eq{model} is equivalent to the S=1/2
ferromagnetic XXZ model in a z-direction magnetic field \eq H =
-J_{xy}\sum_{<ij>}(S_i^xS_j^x+S_i^yS_j^y) - J_z
\sum_{<ij>}S_i^zS_j^z - h_z\sum_i S_i^z.\label{xxz}\ee The mapping
between these two models relates $J_{xy}$ to $t$, $J_z$ to $V$,
$h_z$ to $\mu+Vc/2$ ($c=$ the coordination number). The $z$
component of the magnetization is related to the boson density
according to $M_z=(\nu-1/2)N$. (Here $N$ is the total number of
lattice sites.) For $J_z<J_{xy}$ (i.e.$V<t$), this model has XY
order in the absence of $h_z$. In this range of parameters varying
$h_z$ causes the magnetization $M_z$ to vary continuously, but the
ground state is uniform and, so long as $|M_z|<N/2$, has a
non-zero component of the magnetization which lies in the XY plane
and spontaneously breaks the XY symmetry.  In terms of bosons this
means that the insulating state is Type II, and the doped system
is a uniform superfluid for all $1>\nu>0$.

Conversely, for $J_z>J_{xy}$ ({\it i.e.} $V>t$) the model is effectively
an Ising ferromagnet with a fully polarized  ground state. In
this case $M_z$ exhibits a discontinuity $\Delta M_z=N$ as $h_z$
is varied through zero. If we consider a constrained groundstate,
with $|M_z|<N/2$, the state exhibits phase separation into two
oppositely polarized domains. In terms of the bosons, these two
domains are Mott insulating ($\nu=1$) and empty ($\nu=0$)
respectively. On the boundary between the two behaviors
($J_z=J_{xy}$), the system is equivalent to a Heisenberg
ferromagnet in a magnetic field.  In the absence of $h_z$ the
ground state is N+1 fold degenerate correspond to
$M_z=-N/2,...,N/2$. In the presence of $h_z$ the state with
$M_z=\pm N/2$ ($\pm$ depends on the sign of $h_z$) has the lowest
energy. For this system there is again a first order transition at
$h_z=0$. However, precisely at the transition point the boson
inverse compressibility  $\kappa^{-1}$ diverges, and a uniform
ground state exists for any $\nu$;  at this critical point, only, the
system is neither truly Type I nor Type II.

\section{Duality Transformation for the Bose Mott Insulator in D=2}

There
is a particularly convenient way of thinking about Bose Mott insulators
that  is specific to $D=2$. It turns out that for the class of
model given by \Eq{model},
there is an exact mapping, the ``duality'' transformation, that
provides us an alternative view of the physics of \Eq{model} in
terms of the vortices of the boson field\cite{leesh,fl}. It is
this mapping that enables us to establish a precise connection
between the two types of Mott insulators with Type I and Type II
superconductors\cite{fl,leesh,bfn}.  In the following we discuss
the physical content of the duality transformation without going
into its technical details of the transformation\cite{leesh}.

A vortex is a topological defect in the Bose field. When a boson
is adiabatically transported around a vortex, the boson
wavefunction acquires a phase factor - the Aharonov-Bohm phase
$\theta=2\pi$. In the dual picture, the role of the vortices and
the particles are interchanged:  the vortices are the dual
particles (they turn out to have  Bose statistics as well), and
when a dual particle is transported around an original boson, the
wave  function acquires the same Aharonov-Bohm
phase\cite{leesh,fl} discussed above. The fact that  a boson and a
vortex  acquire a phase when they go around one another implies
that bosons and vortices can not Bose condense simultaneously. As
is well known, in the Bose superfluid phase the vortices must be
absent or localized, {\it i.e.} form an Abrikosov lattice.
Conversely, in the dual phase, where the vortices form a
superfluid\cite{note10}, the boson density (and hence the dual
magnetic flux) must be frozen;  the vortex superfluid phase is the
Mott insulating phase of the original bosons\cite{leesh,fl}.

It is important to note that the absence of dynamical boson density
fluctuation is a necessary but not a sufficient condition for vortex
condensation.
A {\it static} boson density acts like a
background magnetic field to the vortices, which can still
frustrate vortex condensation. However when the static boson
density corresponds to an integral $\nu$, the vortices see a
background magnetic flux corresponding to integral number of flux
quanta per plaquette. (The vortices live on the dual lattice,
i.e., the centers of the square plaquettes.) This type of flux is
``invisible''
because it can be ``gauged away''.
The $\nu=1$ Bose Mott insulator, discussed in the previous
section, corresponds to precisely this situation.

When the boson density  is a fraction ($\nu=p/q$) it is also
possible for the vortices to condense. This is well known\cite{xy} in the
context of the classical frustrated XY model (the XY order
parameter is the Bose amplitude of the vortices) in
$D=2$. Quantum analogues of this, especially for low
order rationals ($q$ a small integer) can readily be imagined.
Such a state is most naturally accompanied by spontaneous
translation symmetry breaking\cite{leesh,oshikawa}, as discussed
above (and as occurs in the classical model), leading to an
enlarged unit cell with an effective integer $\nu$.  However, it
is possible to imagine a more exotic Mott state at $\nu=p/q$ in
which the translation symmetry is unbroken. As has been discussed in Ref. \cite{bfg},
this
could happen if
$q$ elementary vortices form a bound-state, and these composites
then condense. Such a composite condensation is
unfrustrated by an uniform static boson density with $\nu=p/q$.
Moreover, since the condensate vortex consists of $q$ elementary
vortices, charge $1/q$ bosonic soliton excitation (viewed by
the condensate vortices as a flux quantum) can become a finite
energy excitation. Thus there is fractional charge solitons!



In short, a vortex condensate requires the bosons to Mott
insulate. Consequently we can view a boson Mott insulator as a
vortex superconductor\cite{leesh,fl}.  Doping changes the average
background boson density. To the vortices this appears as a change
in the background magnetic field.
The zero temperature boson Mott insulator  is
mapped onto a zero-temperature vortex superconductor (with quantum
fluctuating two-dimensional electromagnetic fields)\cite{note3}.
However, the quantum partition function
of a (particle-hole symmetric)\cite{note10} two dimensional
superconductor with a fluctuating gauge field is equivalent to the
classical ({\it i.e.} thermal) partition function of a three-dimensional
superconductor with thermally fluctuating magnetic
field\cite{note4}.  This leads to the final correspondence between
the Mott insulator and the classical fluctuating 3D superconductor
summarized in Table I.




The properties of classical 3D superconductors in a magnetic field are generally well
known, so most of the comparisons in Table I are self-evident.
We therefore confine ourselves to commenting on a few of the subtleties of this
comparison:

Although a Type I supercondunductor typically expels any applied
field of magnitude $H<H_c$, by choosing an appropriate
experimental geometry it is possible to study its behavior at
fixed  average magnetic induction - which is analogous to studying
the Mott insulator at fixed doping concentration.  This can be
done, for example, by subjecting a flat slab of a Type I
superconductor to a perpendicular magnetic field of strength
$H<H_c$. It is known that when that is done the ``intermediate
state'' consists of a mixture in particular,  among many possible
inhomogeneous structures which have been observed\cite{tink}, one
of the most common is a laminar (or stripe) structure.

Much recent attention has been focussed on fluctuation effects in the  mixed state of a
Type II superconductor. Mean-field theory predicts that
the magnetic induction $B$ increases from zero continuously as  $H$ is raised above the
lower critical field $H_{c1}$. Moreover,  the magnetic
induction first appears in the form of flux tubes each enclosing a single
quantum of magnetic flux
which (in the absence of disorder) form a regular (Abrikosov) lattice.
However, thermal meandering of the flux tubes can affect the physics
dramatically near $H_{c1}$ as well as near the mean-field $H_{c2}$.
Near $H_{c1}$, the distance between neighboring flux
tubes is much greater than the range of their interaction (the
London penetration depth $\lambda$). As a result thermal
meandering of the flux tubes will melt the flux lattice to form a flux
liquid\cite{melt}. At larger magnetic fields the density of flux
tubes becomes higher so that their interaction can stabilize the
flux lattice. This flux lattice persists until $H\ra H_{c2}$ where
the thermal meandering melt the flux lattice again.
The corresponding behavior of a Type II Bose Mott
insulator is a reentrant series of transitions to a superfluid, an insulating crystal of
doped holes, and again a superfluid as a function of increasing doping concentration.

Finally, it is worth recalling that what determines whether a
superconductor is Type I or Type II is the ratio between the
London penetration depth and the core size of the vortices, or
more physically the sign of the interface energy between normal
and superconducting regions\cite{tink}. Type I superconductors
have a positive interface energy while in Type II superconductors
it is negative. As a result, the flux tubes effectively attract
each other in Type I superconductors and repel each other in Type
II. As we have seen in \Eq{model}, this is exactly how we turn a
Type I Mott insulator into a Type II.

\section{Doping a Mott insulator with an order parameter}

As we discussed at the beginning of this paper a Mott insulating
state can be accompanied by translation (and/or other) symmetry breaking. For
example let us consider \Eq{model} with $U/t\rightarrow \infty$
and $V_{ij}=+V$ for nearest neighbor $<ij>$ and $0$ otherwise. For
sufficiently strong $V$ and $\mu=0$ the ground state breaks
translation symmetry and bosons form a checkerboard lattice and
Mott insulate. In this two-fold degenerate ground state the unit
cell is doubled. Is this Mott insulator Type I or Type II?

With the above specific choice of $U$ and $V_{ij}$ \Eq{model} is
equivalent to \Eq{xxz} with $J_{xy}=-t$, $J_z=V$ and $h=\mu-Vc/2$.
For this choice of parameters it is known that as a function of
$h_z$ \Eq{xxz} exhibits a ``spin flop'' transition from the
antiferromagnetic Ising ($S_z$) ordered phase into the
ferromagnetic XY ordered phase. Translate this into the boson
language it implies that
it is Type I.

There is, in fact, a general reason to expect broken symmetry to
increase the tendency of a Mott insulator to be Type I. If there
is a broken symmetry in the insulating state, then there generally
is a corresponding local
order parameter ($\psi$)  (for the above example
$\psi$ is the two-sublattice density wave order parameter).
We now consider the effect of this order parameter on the interface energy between the
insulating and the metallic (doped) state.
Typically,
$\psi$ will have a reduced value (perhaps 0) in the doped state.
Thus, at the
interface between doped and undoped region a spatially varying
$\psi$ is necessary. The spatial gradient energy \eq \int
d^dx |\gr\psi|^2\ee  will thus make a positive contribution to the interface energy,
 increasing the tendency of the Mott insulator to be   Type I.

\section{An example of translationally invariant Bose Mott insulator at $\nu=1/2$}

A recently discovered system which exhibits a Bose Mott insulating
state at a fractional $\nu$ is the quantum hardcore dimer model on
a triangular lattice.  The model is defined in terms of bosons
(called dimers) which live on the nearest neighbor bonds of a
lattice, {\it i.e.} $b_{ij}^{\dagger}$ creates a boson on the bond
connecting sites i and j.  We impose a hardcore constraint - this
can be thought of as deriving from the $U\to \infty$ limit of a
contact interaction, which forbids two dimers to occupy the bonds
emanating from a single site, {\it i.e.}
$b^{\dagger}_{ij}b_{ij}b^{\dagger}_{jk}b_{jk}=0$ for any triplet
of nearest neighbor sites, $<ijk>$.  Thus, since each bond touches
exactly two sites, it is possible
on any regular
lattice with one site per unit cell to satisfy this constraint if and only if $\nu \le
1/2$. The simplest (shortest range) Hamiltonian\cite{rk} one can construct
for these dimers is
\eqa H= &-t\sum_{(ijkl)}[
b^{\dagger}_{ij}b^{\dagger}_{kl}b_{il}b_{jk}+ {\rm h.c.}] \\ &+V
\sum_{(ijkl)}[ b^{\dagger}_{ij}b^{\dagger}_{kl}b_{kl}b_{ij}+
b^{\dagger}_{il}b^{\dagger}_{kj}b_{kj}b_{il}] \nonumber \eea
where
$(ijkl)$ denotes a ``square'' of nearest-neighbor sites, such that
$<ij>$, $<jk>$, $<kl>$, and $<li>$ are all pairs of
nearest-neighbors.

On a square lattice, all zero temperature Mott insulating phases with
$\nu=1/2$ derived from this Hamiltonian are believed to break
translational symmetry, at least doubling the unit cell size, so that the
effective
$\nu$ is integral\cite{sondhi}.  However, on a triangular lattice, it has now been
established\cite{sondhi} that for a range of parameters near
$t\sim V$, there is a dimer liquid phase, with exponentially falling
correlation functions and no broken symmetries.  From the dual viewpoint,
this phase can be viewed as a superfluid of vortex pairs\cite{bfg}.  As mentioned
previously, there are observable topological consequences of the vortex
pairing, including certain predictable ground-state degeneracies on
closed surfaces.

This same model can serve as a paradigmatic example of a Mott
insulating state of spinfull fermions at an odd integer $\nu=1$.  To see
this, we merely need to define fermion creation operators,
$c^{\dagger}_{\sigma,ij}$, for an electron with spin polarization
$\sigma$ on bond $<ij>$.  In terms of this, we can express the dimer
creation operators as
\eq
b^{\dagger}_{ij}=c^{\dagger}_{\uparrow,ij}c^{\dagger}_{\downarrow,ij}
\ee
and an additional constraint,
$\sum_{\sigma}c^{\dagger}_{\sigma,ij}c^{\dagger}_{\sigma,ij}=0$ or 2,
which can be thought of as a consequence of a strong attractive interaction
between two electrons on the same bond.

Is this sort of fractional Mott insulator Type I or Type II?  The intensive study of what
happens when one dopes such a ``spin-liquid'' state was
initiated by the proposal\cite{pwa} that doping such a  liquid would lead inevitably
to high temperature superconductivity.  The basis for this proposal was
the thought that pairing correlations ({\it e.g.} a spingap\cite{krs})
might be present already in the insulating state, and these  would evolve
smoothly into superconducting pairing upon doping.  Central to this
proposal is the assumption that the spin-liquid is a Type II Mott
insulator.  Indeed, an early triumph of this idea was the observation by
Ioffe and Larkin\cite{il,rk} that, for at least a range of parameters,
doped holes in the quantum dimer model do not phase separate.  Moreover,
reversing the logic of the previous section, the fact that  this state
has no broken symmetry increases the likelihood that it is Type II.

\section{conclusions}

To summarize, in this work we introduce the concept that there are
two types of Mott insulator. In particular, there exist a whole
class of Mott insulators, the type I Mott insulators, that become
inhomogeneous upon doping. The type I behavior does not have to
originate from order parameter competition. We give a familiar
lattice boson example where the Mott state exhibits no symmetry
breaking, however depending on the sign of certain interaction, it
can be either type I or type II after doping. Through duality
these two types of insulating states have a close tie to Type I
and Type II superconducting states.\cite{bfn} Finally after addressing the
``zeroth order'' physics of doped Mott insulator, we point out
that many important ``first order'' issues remain unanswered.
Among them are ``what determines the structure of the
inhomogeneity in a doped Type I Mott insulator?'' ``What is the
ultimate low energy behavior of these inhomogeneities?''
\\

Acknowledgements:  We wish to thank C.Nayak for useful discussions.  This work was
supported in part by NSF grants \# DMR 99-71503 (DHL) and DMR - 0110329 (SAK).

\widetext


\begin{thebibliography}{99}
\bibitem{band} A band insulator insulates because of
Pauli exclusion principle. In the non-interacting prototype of
this type of insulator all available single-particle orbitals are
occupied. This type of insulator usually possesses a even number
of electrons per crystalline unit cell.
\bibitem{note} A Mott insulator necessarily has a charge gap, however it is
possible for it to have  gapless neutral excitations.
\bibitem{coulomb} Strictly speaking, in the limit  $\delta\equiv\nu^*-\nu\ra 0$,
since the Madulung energy per hole diverges, the groundstate in
the presence of long-range Coulomb interactions is  always a
Wigner crystal of holes, independently of whether in the absence of such
interactions the state would be Type I or Type II. However, where
the Coulomb interactions are weak, due to the presence of a large
dielectric constant,     clustering occurs in a Type I Mott
insulator for  $\delta>\delta_c$, where $\delta_c$ is a
parametrically small (in powers of  $\epsilon^{-1}$  ) critical
doping concentration.  In the absence of long-range forces and for
small enough $\delta$ , the charges in a Type II Mott insulator
always form a uniform fluid.
\bibitem{noteaf}  There are, in each case, various additional conditions concerning the
nature of the antiferromagnet specified in the various conjectural
relations between antiferromagnetism and Type I behavior.  Before
gleefully constructing proposed counterexamples, such as a doped
Ising antiferromagnet with additional strong, short-range
repulsive interactions, the reader should consult the original
papers.
\bibitem{dope} The system considered in this Ref.\cite{bosemott}
can  be ``doped'' by varying the period of the trap lattice.
\bibitem{note10} We note that the vortex superfluid discussed
here differs from the generic boson superfluid because the net
vorticity is zero. The field theory that describes such
"particle-hole symmetric" condensate has space and time on equal
footing (i.e. relativistic). This fact makes the later analogy
with thermal fluctuating 3D superconductors completely
appropriate.
\bibitem{note3} The field strength $F_{\mu\nu}$ ($\mu=0,1,2$) of a 2+1-dimensional EM field has three independent components.
The correspondence of the these three components with the boson three-current $(\rho,\v j)$ is as follows:
$F_{12}\leftrightarrow \rho, F_{01}\leftrightarrow j_2, F_{02}\leftrightarrow j_1$.
\bibitem{note4} In the correspondence between a quantum superconductor at $T=0$ and a thermally fluctuating
superconductor at finite temperature the magnetic field ($F_{12}$) and electric field
($F_{01},F_{02}$) in the 2+1 dimensional quantum theory are mapped on to $B_z$ and ($B_y,
-B_x$) respectively.
\end{thebibliography}
\end{document}